\documentclass[aps,prl,reprint,superscriptaddress]{revtex4-1}
\usepackage{amsmath}
\usepackage{amsfonts}
\usepackage{amssymb}
\usepackage{amsthm}
\usepackage{graphicx}
\usepackage{CJK}
\usepackage{color}
\usepackage{times}
\usepackage{mathptmx}
\usepackage[colorlinks, linkcolor=blue, urlcolor=blue, anchorcolor=blue, citecolor=blue]{hyperref}
\usepackage{mathtools}

\begin{document}


\title{Stabilizing current-driven steady flows of 180$^{\circ}$ domain walls in spin valves by interfacial Dzyaloshinskii-Moriya interaction}


\author{Jiaxin Du}
\affiliation{College of Physics and Hebei Advanced Thin Films Laboratory, Hebei Normal University, Shijiazhuang 050024, People’s Republic of China}
\author{Mei Li}
\email{limeijim@163.com}
\affiliation{Physics Department, Shijiazhuang University, Shijiazhuang, Hebei 050035, People’s Republic of China}
\author{Jie Lu}
\email{jlu@hebtu.edu.cn}
\affiliation{College of Physics and Hebei Advanced Thin Films Laboratory, Hebei Normal University, Shijiazhuang 050024, People’s Republic of China}


\date{\today}

\begin{abstract}
Modulations of interfacial Dzyaloshinskii-Moriya interaction (iDMI) on current-driven
dynamics of 180$^{\circ}$ domain walls (180DWs) in long and narrow spin valves (LNSVs) 
with heavy-metal caplayers are systematically investigated.
We focus on LNSVs with in-plane magnetic anisotropy in their free layers.
For planar-transverse polarizers (pinned layers of LNSVs), the Walker breakdown can be
postponed considerably (practical infinity) by iDMI.
More interestingly, the originally unstable traveling mode is also stabilized 
by iDMI with high saturation velocity thus serves as fast carrier of information. 
For parallel polarizers, the Walker limit is increased and the corresponding modifications of iDMI to wall velocity in both steady and precessional flows of 180DWs are provided.
For perpendicular polarizers, precessional flow of 180DWs is absent due to the stability 
of stationary mode beyond the modified Walker limit.
For LNSVs with perpendicular magnetic anisotropy in their free layers, similar results
are obtained.
Our findings open new possibilities for developing magnetic nanodevices based on 180DW propagation with low energy consumption and high robustness.
\end{abstract}


\maketitle

\section{\label{Section_introduction} I. Introduction} 
Spin valves, as the first-proposed host system of spin-transfer torque (STT)\cite{Slonczewski_JMMM_1996}, have attracted
considerable attention in the past decades for both academic and industrial interests\cite{Fert_JAP_2002,Fert_APL_2003,Lim_APL_2004,Rebei_Mryasov_PRB_2006,Kawabata_IEEE_2011,Khvalkovskiy_PRL_2009,Boone_PRL_2010_exp,Grollier_NatPhys_2011,Metaxas_SciRep_2013,Grollier_APL_2013,jlu_PRB_2019,He_EPJB_2013}.
180$^{\circ}$ domain walls (180DWs) in their free layers are driven to move 
along the long axis by perpendicularly injected currents [filtered by pinned layers (polarizers)
first and become spin-polarized],
thus realizes the transition between states with high and low resistances.
To acquire high performance of nanodevices based on spin valves, it is crucial to 
achieve high enough velocity of 180DWs for a given current density.
Most existing works focus on the cases where free layers of spin valves
have in-plane magnetic anisotropy (IPMA).
Early simulations on parallel and perpendicular polarizers only considered 
the Slonczewski torque (SLT) and turn out that
current densities of several $10^{8}\ \mathrm{A/cm^2}$ only induce 180DW velocity around $100$ m/s\cite{Rebei_Mryasov_PRB_2006,Kawabata_IEEE_2011}.
In 2009, Khvalkovskiy \textit{et. al.} considered both the SLT and the field-like torque (FLT)
and revealed that to achieve a wall velocity of 100 m/s, 
the current density for parallel polarizers is lowered to $10^{7}\ \mathrm{A/cm^2}$,
while for perpendicular polarizers, the current density is further decreased to $10^{6}\ \mathrm{A/cm^2}$\cite{Khvalkovskiy_PRL_2009}.
Later, these numerical results were confirmed by transport measurements 
in long and narrow spin valves (LNSVs)\cite{Boone_PRL_2010_exp} and half-ring MTJs\cite{Grollier_NatPhys_2011,Metaxas_SciRep_2013,Grollier_APL_2013}.

In 2019, we revisited this issue in LNSVs
with planar-transverse polarizers and obtained several interesting findings\cite{jlu_PRB_2019}: 
(i) When the STT coefficient is assumed to be constant (the simplest way to deal with it),
all the steady flows with finite velocity are unstable thus not valuable in application.
(ii) When the STT coefficient takes the Slonczewski's original form,
stable steady flow of 180DWs with finite velocity can 
survive for strong enough planar-transverse polarizers.
Meantime the current efficiency is comparable with that of perpendicular ones. 
(iii) More importantly, 180DWs have ultrahigh differential mobility around the 
onset of stable wall excitation. 
Based on these results, magnetic nanodevices with low energy consumption and high sensitivity,
for example the magnetic nanoswitches, can be proposed.
However, the success of these devices depends on strong enough polarization effect
which needs thick pinned layers thus deviates from the trend of miniaturization.

In this work, within the Lagrangian framework\cite{jlu_PRB_2019,He_EPJB_2013,Boulle_PRL_2013}
we illuminate that the interfacial Dzyaloshinskii-Moriya interaction (iDMI)\cite{Dzyaloshinsky,Moriya}
from an extra heavy-metal caplayer over the free layer can stabilize
the steady flows of 180DWs in the free layer under planar-transverse polarizers,
and further lead to the ``practical absence of Walker breakdown".
In addition, the wall velocity saturates to an iDMI-determined high value.
These interesting behaviors provide higher redundancy and robustness of in-plane oriented polarizers
for highest current efficiency without requiring that it must be parallel to the easy axis.
However, at the same time we lose the ``ultrahigh differential mobility" behavior at a cost.
Also, the modulations of iDMI to Walker limit, as well as wall velocity in both steady and 
precessional flows under parallel and perpendicular polarizers are provided.

\section{\label{Section_Lagrangian} II. Model and method}
\begin{figure} [htbp]
	\centering
	\includegraphics[width=0.43\textwidth]{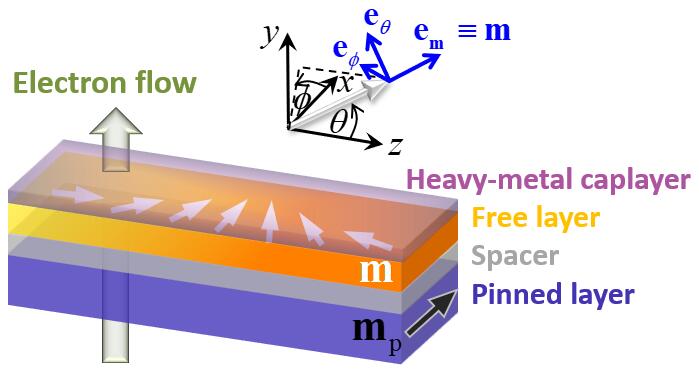}
	\caption{(Color online) Sketch of a LNSV with a heavy-metal caplayer.
		The LNSV has a three-layer structure: a pinned FM layer ($\mathbf{m}_{\mathrm{p}}$, polarizer), a NM metallic spacer and a free FM layer ($\mathbf{m}$).
		A 180DW in the free layer is driven to move along the long axis of LNSV by 
		perpendicularly injected current $J_{\mathrm{charge}}=-J_e\mathbf{e}_y$.
		($\mathbf{e}_x,\mathbf{e}_y,\mathbf{e}_z$) is the global Cartesian coordinate system, and
		($\mathbf{m},\mathbf{e}_{\theta},\mathbf{e}_{\phi}$) forms the local spherical coordinate
		system associated with $\mathbf{m}$.}\label{fig1}
\end{figure}

A LNSV with a heavy-metal caplayer is depicted in Fig. \ref{fig1}.
The original three-layer structure of the LNSV is:
a free ferromagnetic (FM) layer with tunable magnetization texture,
a nonmagnetic (NM) metallic spacer and a pinned FM layer with a fixed magnetization orientation (polarizer).
In the global Cartesian coordinate system,
$\mathbf{e}_z$ is along the long axis of LNSV, $\mathbf{e}_y$ follows the plane normal and $\mathbf{e}_x=\mathbf{e}_y\times\mathbf{e}_z$.
In the main text, we consider free layers with IPMA and easy axis being along $\mathbf{e}_z$. 
The results for perpendicular magnetic anisotropy (PMA) are quite similar so we only briefly summarize them in the discussion section.
On the other hand, the polarizers are usually made of hard FM materials. 
Its magnetization orientation ($\mathbf{m}_{\mathrm{p}}$) 
has three typical choices:
(a) $\mathbf{m}_{\mathrm{p}}=\mathbf{e}_z$ (parallel),
(b) $\mathbf{m}_{\mathrm{p}}=\mathbf{e}_y$ (perpendicular) and
(c) $\mathbf{m}_{\mathrm{p}}=\mathbf{e}_x$ (planar-transverse).
At last, the charge current is $J_{\mathrm{charge}}=-J_e\mathbf{e}_y$ with $J_e>0$ being the electron density,
which ensures that electrons flow from the pinned to free layers.

The total magnetic energy density $\mathcal{E}$ of the FM free layer includes
the exchange, crystalline anisotropy, magnetostatics, FLT-induced effective potential\cite{jlu_PRB_2019,He_EPJB_2013,Boulle_PRL_2013}
and the iDMI contribution\cite{Bogdanov_JMMM_1994}:
\begin{equation}\label{magnetic_energy_density_total}
\begin{split}
\mathcal{E}[\mathbf{m}]& = A\left(\nabla\mathbf{m}\right)^2+\mu_0 M_s^2\left(-\frac{1}{2}k_{\mathrm{E}}m_z^2+\frac{1}{2}k_{\mathrm{H}}m_y^2\right)   \\
&\quad -\mu_0 M_s \xi a_J p_{\mathbf{m}} + D_{\mathrm{i}}\left[m_y\nabla\cdot\mathbf{m}-\left(\mathbf{m}\cdot\nabla\right)m_y \right],
\end{split}
\end{equation}
where $\mu_0$ is the vacuum permeability, $A$, $M_s$ and $\mathbf{m}(\mathbf{r})$ are the exchange stiffness, 
saturation magnetization, and unit magnetization vector field of the free layer, respectively.
Far from the ends or corners of LNSVs, most of the magnetostatic energy can be described by local
quadratic terms of $M_{x,y,z}$ by means of three average demagnetization factors $D_{x,y,z}$\cite{jlu_PRB_2019,Aharoni_JAP_1998,jlu_PRB_2016,jlu_SciRep_2017,jlu_PRB_2020,jlu_JMMM_2021},
thus in Eq. (\ref{magnetic_energy_density_total}) $k_{\mathrm{E}}(k_{\mathrm{H}})$ denotes 
the total anisotropy coefficient along the easy (hard) axis of free layer, 
namely $k_{\mathrm{E}}=k_1 + (D_x-D_z)$ and $k_{\mathrm{H}}=k_2 + (D_y-D_x)$
with $k_{1(2)}$ being the crystalline anisotropy coefficient in easy (hard) axis.
The SLT strength reads $a_J=g(P,\mathbf{m},\mathbf{m}_{\mathrm{p}})\hbar J_e/(\mu_0 g_e d e M_s)$,
in which $\hbar$ is the Planck constant, $g_e$ is the electron g-factor,
$d$ is the thickness of free layer, $e(>0)$ is the absolute charge of electron,
and $P$ is the spin polarization of the current filtered by the pinned layer.
$g(P,\mathbf{m},\mathbf{m}_{\mathrm{p}})$ is the dimensionless spin polarization factor, 
and in the simplest case, $g(P,\mathbf{m},\mathbf{m}_{\mathrm{p}})\equiv P$\cite{Khvalkovskiy_PRL_2009,Lee_PhysRep_2013,Chshiev_PRB_2015}.
$\xi$ describes the relative strength of FLT over SLT.
In addition, $\mathbf{m}$ is fully described by its polar and azimuthal angles $(\theta,\phi)$,
leading to the associated local spherical coordinate system ($\mathbf{e}_{\mathbf{m}}\equiv \mathbf{m},\mathbf{e}_{\theta},\mathbf{e}_{\phi}$).
Then $\mathbf{m}_{\mathrm{p}}$ is decomposed into
$\mathbf{m}_{\mathrm{p}}=p_{\mathbf{m}}\mathbf{e}_{\mathbf{m}}+p_{\theta}\mathbf{e}_{\theta}+p_{\phi}\mathbf{e}_{\phi}$
with
\begin{equation}\label{mp_expression_in_sperical_definitions}
\begin{split}
p_{\mathbf{m}}&=\sin\theta_{\mathrm{p}}\cos(\phi-\phi_{\mathrm{p}})\sin\theta+\cos\theta_{\mathrm{p}}\cos\theta,   \\
p_{\theta}&=\sin\theta_{\mathrm{p}}\cos(\phi-\phi_{\mathrm{p}})\cos\theta-\cos\theta_{\mathrm{p}}\sin\theta,   \\
p_{\phi}&=-\sin\theta_{\mathrm{p}}\sin(\phi-\phi_{\mathrm{p}}),
\end{split}
\end{equation}
where $\theta_{\mathrm{p}}$ ($\phi_{\mathrm{p}}$) is the polar (azimuthal) angle of $\mathbf{m}_{\mathrm{p}}$
in global ($\mathbf{e}_x,\mathbf{e}_y,\mathbf{e}_z$) coordinate system.
At last, $D_{\mathrm{i}}$ is the iDMI strength.

The dynamics of $\mathbf{m}(\mathbf{r})$ is described by the Lagrangian\cite{jlu_PRB_2019,He_EPJB_2013,Boulle_PRL_2013}
\begin{equation}\label{Lagrangian_density}
\mathcal{L}=\frac{\mu_0 M_s}{\gamma_0}\dot{\phi}\cdot (1-\cos\theta)-\mathcal{E},
\end{equation}
where $\gamma_0=\mu_0\gamma_e$ with $\gamma_e$ being the gyromagnetic ratio of electrons 
and an overdot means $\partial/\partial t$.
To include various damping processes,
an extra dissipation density is introduced as 
\begin{equation}\label{Dissipation_density}
\frac{\mathcal{F}}{\mu_0 M_s^2}=\frac{\alpha}{2}\frac{\dot{\theta}^2+\dot{\phi}^2\sin^2\theta}{\gamma_0 M_s}- \frac{a_J}{M_s} (p_{\theta}\sin\theta\dot{\phi}-p_{\phi}\dot{\theta}).
\end{equation}
The corresponding Lagrangian-Rayleigh equation
\begin{equation}\label{Eular_Lagrangian_equation}
\frac{\mathrm{d}}{\mathrm{d}t}\left(\frac{\delta\mathcal{L}}{\delta\dot{X}}\right)-\frac{\delta\mathcal{L}}{\delta X}+\frac{\delta\mathcal{F}}{\delta \dot{X}}=0
\end{equation}
provides full description of magnetization dynamics, where $X$ represents any related generalized coordinates.

Next, the configuration space of 180DWs is expanded by the generalized Walker ansatz\cite{Walker_JAP_1974}
\begin{equation}\label{Walker_static_generalized}
\ln\tan\frac{\vartheta(z,t)}{2}=\eta\frac{z-q(t)}{\Delta(t)},\quad \phi(z,t)\equiv\varphi(t),
\end{equation}
with the three collective coordinates $q(t)$, $\varphi(t)$ and $\Delta(t)$ indicating 
the wall center position, tilting angle and width, respectively.
Here $\eta=+1$ or $-1$ represents head-to-head or tail-to-tail 180DWs.
By letting $X$ take $q(t)$, $\varphi(t)$, $\Delta(t)$ successively,
and integrating over the long axis (i.e. $\int_{-\infty}^{+\infty}\mathrm{d}z$), 
Eq. (\ref{Eular_Lagrangian_equation}) provides the following
dynamic equations
\begin{subequations}\label{Dynamical_equation_original}
	\begin{align}
	\alpha\eta\frac{\dot{q}}{\Delta}+\dot{\varphi}&=\gamma_0\left(\frac{\pi}{2}a_J p_{\varphi}+\xi a_J \cos\theta_{\mathrm{p}}\right), \\
	\eta\frac{\dot{q}}{\Delta}-\alpha\dot{\varphi}&=\gamma_0 M_s k_{\mathrm{H}}\sin\varphi\cos\varphi
	-\frac{\eta D_{\mathrm{i}}\gamma_0\pi}{2\mu_0 M_s\Delta}\cos\varphi   \nonumber \\
	& \qquad +\gamma_0\left(a_J \cos\theta_{\mathrm{p}}-\frac{\pi}{2}\xi a_J p_{\varphi}\right), \\
	\frac{\pi^2\alpha}{6}\frac{\dot{\Delta}}{\Delta}&=\gamma_0 M_s\left(\frac{l_0^2}{\Delta^2}-k_{\mathrm{E}}-k_{\mathrm{H}}\sin^2\varphi\right)   \nonumber  \\
	& \qquad +\gamma_0\pi \xi a_J \sin\theta_{\mathrm{p}}\cos(\varphi-\phi_{\mathrm{p}}),
	\end{align}
\end{subequations}
with $l_0\equiv\sqrt{2A/(\mu_0 M_s^2)}$ being the exchange length of the free layer.
The emergence of iDMI term leads to all the interesting results in this work,
which will be introduced in the following sections.

\section{III. Walker breakdown postponement and stable steady flows under planar-transverse polarizers}
For planar-transverse polarizers, $\mathbf{m}_{\mathrm{p}}=\mathbf{e}_x$ thus $\theta_{\mathrm{p}}=\pi/2$ 
and $\phi_{\mathrm{p}}=k\pi$. The dynamical equations turn to
\begin{subequations}\label{Dynamical_equation_pt_polarizer}
	\begin{align}
	\frac{1+\alpha^2}{\gamma_0 k_{\mathrm{H}}M_s}\frac{\eta\dot{q}}{\Delta}&=\sin\varphi\left[\cos\varphi+\frac{\pi}{2}\frac{(-1)^k(\xi-\alpha)a_J}{k_{\mathrm{H}}M_s}\right]-\frac{\lambda}{2}\cos\varphi,   \\
	\frac{1+\alpha^2}{\alpha\gamma_0 k_{\mathrm{H}}M_s}\dot{\varphi}&=\frac{\lambda}{2}\cos\varphi -\sin\varphi\left[\cos\varphi+\frac{\pi}{2}\frac{(-1)^k(1+\alpha\xi)a_J}{\alpha k_{\mathrm{H}}M_s}\right],  \\
	\frac{\pi^2\alpha}{6\gamma_0 M_s}\frac{\dot{\Delta}}{\Delta}&=\left(\frac{l_0^2}{\Delta^2}-k_{\mathrm{E}}-k_{\mathrm{H}}\sin^2\varphi\right)+\frac{\pi\xi a_J}{M_s}(-1)^k\cos\varphi,
	\end{align}
\end{subequations}
in which $\lambda\equiv \eta D_{\mathrm{i}}\pi/(k_{\mathrm{H}}\mu_0 M_s^2\Delta)$ denotes
the relative strength of iDMI.
In this section we demonstrate the Walker breakdown postponement
and stabilization effect of iDMI to steady flows of 180DWs, which are the central highlights of this work.

For steady flows of 180DWs, $\dot{\varphi}=\dot{\Delta}=0$, thus one has
\begin{equation}\label{planar_transverse_fx_definition}
f(\varphi,\lambda)\equiv\frac{\lambda\cos\varphi-\sin 2\varphi}{2\sin\varphi}=(-1)^k\frac{J_e}{J_{W,x}^{0}},
\end{equation}
in which $J_{W,x}^{0}\equiv 2\alpha k_{\mathrm{H}}\mu_0 M_s^2 g_e d e/[(1+\alpha\xi)\pi P\hbar]$
is the original Walker limit in the absence of iDMI ($\lambda=0$). 
To simplify our discussion, several symmetrical features of the above equation 
are revealed first (note that  $J_e>0$ always holds).
If $\varphi_0$ is a solution of Eq. (\ref{planar_transverse_fx_definition}) for a given combination of 
$[\lambda,J_e,(-1)^k]$, then:
(i) $-\varphi_0$ must be a solution for $[-\lambda,J_e,(-1)^k]$; 
(ii) $\pi-\varphi_0$ must be a solution for $[\lambda,J_e,(-1)^{k+1}]$. 
Therefore we can focus our discussion on the case where $\lambda>0$, $J_e>0$ and $k=0$,
so that $(-1)^k=+1$.

\begin{figure} [htbp]
	\centering
	\includegraphics[width=0.45\textwidth]{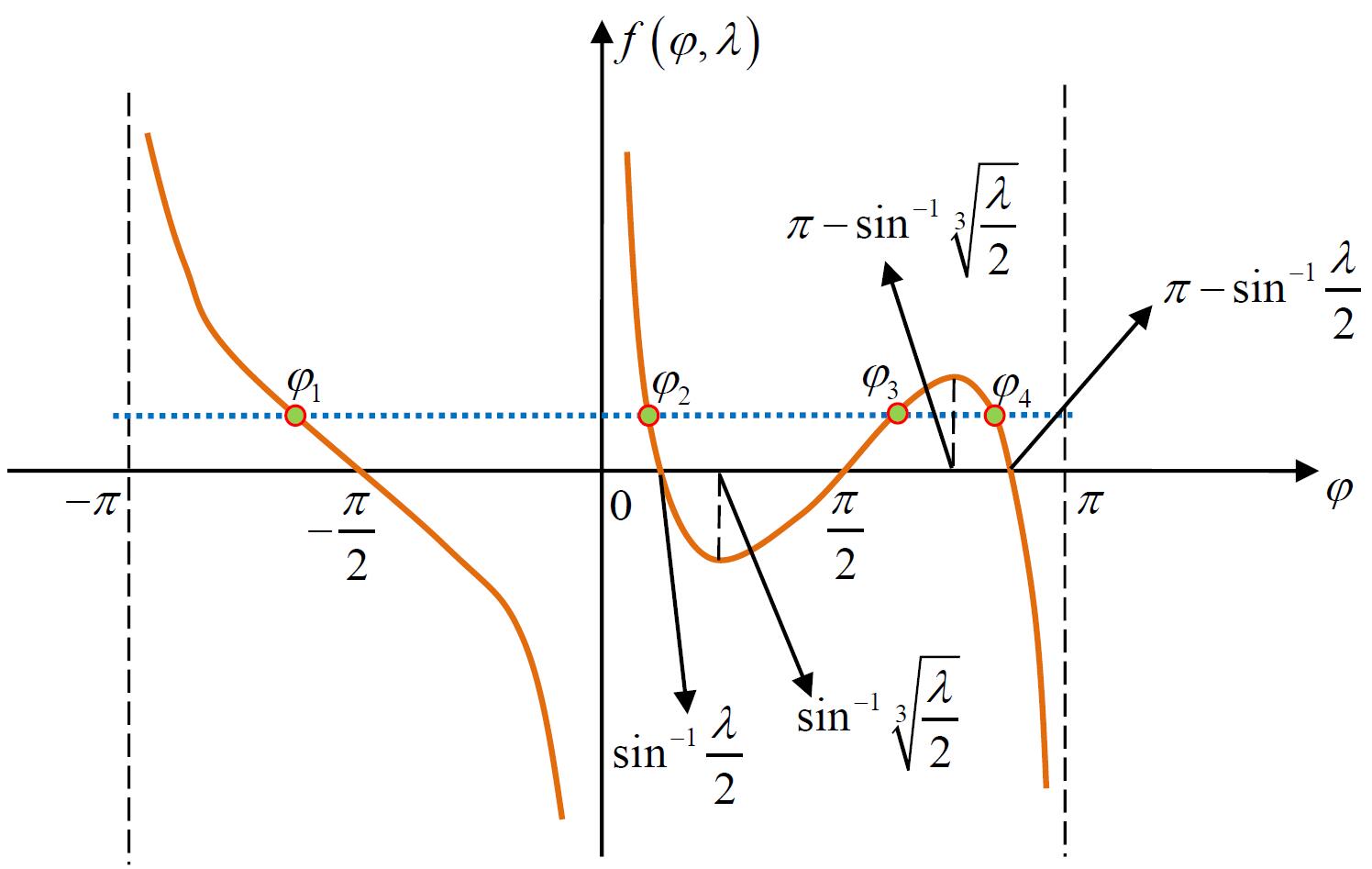}
	\caption{(Color online) Illustration of Eq. (\ref{planar_transverse_fx_definition})
	for $0<\lambda<2$: 
	orange solid curves represent the two branches of function $f(\varphi,\lambda)$
	within  $(-\pi,0)$ and $(0,\pi)$, respectively.
	Blue dotted line indicates a given $J_e$ with $k=0$.
	Their four [$0<J_e/J_{W,x}^{0}<\Gamma(\lambda)$] or two [$J_e/J_{W,x}^{0}>\Gamma(\lambda)$] 
	intersections are the solutions of Eq. (\ref{planar_transverse_fx_definition}). }\label{fig2}
\end{figure}

Now the function $f(\varphi,\lambda)$ has two distinct branches which lie respectively 
in $(-\pi,0)$ and $(0,\pi)$, as shown by orange solid curves in Fig. \ref{fig2}.
The branch in $(-\pi,0)$ is monotonically decreasing, thus provide the first 
solution $\varphi_1$ of Eq. (\ref{planar_transverse_fx_definition}) lying within $(-\pi,-\pi/2)$ for $J_e>0$.
On the other hand, the branch in $(0,\pi)$ is nonmonotonic.
For not too strong iDMI ($|\lambda|<2$), which is the case in most real LNSVs,
it first decreases from $+\infty$ at $\varphi=0$ to 0 at $\varphi=\arcsin(\lambda/2)$
and further to the local minimum $-F(\lambda)$ at $\varphi=\arcsin\sqrt[3]{\lambda/2}$,
then increases to 0 at $\varphi=\pi/2$ and further to the local maximum 
$F(\lambda)$ at $\varphi=\pi-\arcsin\sqrt[3]{\lambda/2}$.
After that, it decreases to 0 at $\varphi=\pi-\arcsin(\lambda/2)$ and finally to $-\infty$ at $\varphi=\pi$.
Here
\begin{equation}\label{planar_transverse_fx_max}
F(\lambda)=\sqrt{3\left(\frac{\lambda}{2}\right)^{\frac{4}{3}}-\left(\frac{\lambda}{2}\right)^2-3\left(\frac{\lambda}{2}\right)^{\frac{2}{3}}+1}.
\end{equation}
Consequently, for any $J_e>0$ there always exists the second solution $\varphi_2$ of 
Eq. (\ref{planar_transverse_fx_definition}) satisfying $0<\varphi_2<\arcsin(\lambda/2)<\arcsin\sqrt[3]{\lambda/2}$.
Other two solutions beyond $\pi/2$ only exist for $0<J_e/J_{W,x}^{0}<F(\lambda)$:
$\pi/2<\varphi_3<\pi-\arcsin\sqrt[3]{\lambda/2}$ and $\pi-\arcsin\sqrt[3]{\lambda/2}<\varphi_4<\pi-\arcsin(\lambda/2)$.

Next we perform stability analysis of the solutions $\varphi_{i}(i=1,2,3,4)$ to Eq. (\ref{planar_transverse_fx_definition}).
Suppose the azimuthal angle $\varphi$ deviates a little bit from a certain exact solution, that is,  $\varphi=\varphi_i+\delta\varphi_i$.
Substituting it into Eq. (\ref{Dynamical_equation_pt_polarizer}b) and after simple algebra, one has
\begin{equation}\label{Stability_analysis_phi_pt_polarizer}
\frac{\partial(\delta\varphi_i)}{\partial t}=\frac{\gamma_0 \alpha M_s k_{\mathrm{H}}}{1+\alpha^2}\left[\sin\varphi\frac{\partial f(\varphi,\lambda)}{\partial \varphi}\right]_{\varphi=\varphi_i}\delta\varphi_i.
\end{equation}
The stability of $\varphi_i-$solution requires the terms in brackets to be negative. 
Obviously, $\varphi_{1,3}$ fail and $\varphi_{2,4}$ succeed.
Generally, $\varphi_4$ provides one possibility of steady flow when 
$0<J_e/J_{W,x}^{0}<\Gamma(\lambda)$.
More interestingly, $\varphi_2$ results in the considerable enhancement of Walker limit,
however the rigorous ``universal absence of Walker breakdown" is forbidden by
the existence condition of wall width from Eq. (\ref{Dynamical_equation_pt_polarizer}c),
which reads
\begin{equation}\label{Stability_analysis_J_e_up}
J_e<J_x^{\mathrm{u}}(\varphi_2)\equiv J_{W,x}^{0}\cdot\frac{1+\alpha\xi}{2\alpha\xi}\cdot\frac{\sin^2\varphi_2+k_{\mathrm{E}}/k_{\mathrm{H}}}{\cos\varphi_2}.
\end{equation}
Since $\varphi_2\rightarrow \phi_{\mathrm{p}}=0$ as $J_e$ increases, $J_x^{\mathrm{u}}(\varphi_2)$ has
the infimum $\tilde{J}_x^{\mathrm{u}}\equiv J_{W,x}^{0}[(1+\alpha\xi)/(2\alpha\xi)](k_{\mathrm{E}}/k_{\mathrm{H}})$.
In typical LNSVs bearing free layers with IPMA, $k_{\mathrm{E}}/k_{\mathrm{H}}\sim 0.1$ to 1, $\alpha\sim 0.01$ and
$\xi\sim 0.1$. 
Consequently, $\tilde{J}_x^{\mathrm{u}}\gg J_{W,x}^{0}$ and is usually out of experimental accessibility.
Therefore, the Walker breakdown can be viewed to be postponed to infinity in practical sense.

At last, after putting the stable tilting angle $\varphi_2$ into Eq. (\ref{Dynamical_equation_pt_polarizer}a),
the wall velocity reads $\dot{q}=-\eta\gamma_0 k_{\mathrm{H}} M_s\Delta(\lambda\cos\varphi_2-\sin 2\varphi_2)/[2(1+\alpha\xi)]$.
It approaches
\begin{equation}\label{planar_transverse_v_saturation}
v_{\mathrm{sat}}=-\frac{\pi\gamma_e D_{\mathrm{i}}}{2(1+\alpha\xi)M_s},
\end{equation}
when $J_e$ increases to several times of $J_{W,x}^{0}$ thus $\varphi_2\approx\phi_{\mathrm{p}}=0$
(since $k=0$). 
Eq. (\ref{planar_transverse_v_saturation}) implies several interesting facts:
(i) the iDMI-stabilized steady flow of 180DWs under planar-transverse polarizers
has a saturation velocity $v_{\mathrm{sat}}$;
(ii) $v_{\mathrm{sat}}$ is independent on the wall's topological charge $\eta$;
(iii) $v_{\mathrm{sat}}$ is even irrelevant to (both crystalline and shape) magnetic anisotropy,
and solely determined by the iDMI strength;
(iv) $v_{\mathrm{sat}}$ changes sign when $\mathbf{m}_{\mathrm{p}}$ rotates from 
$+\mathbf{e}_x$ to $-\mathbf{e}_x$ since $\varphi_2\rightarrow\phi_{\mathrm{p}}$
as the driving current increases.

To acquire a quantitative impression of our analytics, we choose typical magnetic parameters to 
do some numerical estimations:
$D_{\mathrm{i}}=0.5$ $\mathrm{mJ/m^2}$, $M_s=500$ kA/m, $A=1.57\times 10^{-11}$ J/m, $\alpha=0.01$ and $\xi=0.1$.
Thus the exchange length $l_0=10$ nm.
The thickness of free layer is 3 nm and $k_{\mathrm{E}}\approx k_{\mathrm{H}}\approx 1$.
In addition, the polarization of current is assumed to be $P=0.5$.
Based on them, the original Walker limit reads $J_{W,x}^0=3.6\times 10^6$ $\mathrm{A/cm^2}$.
The dimensionless iDMI parameter $|\lambda|=1/2$ which satisfies $|\lambda|<2$.
At last, the saturation velocity is $v_{\mathrm{sat}}\approx -270$ m/s when $J_e$ is several times 
as large as $J_{W,x}^0$.
This means that for current density around $10^6 \sim 10^7$ $\mathrm{A/cm^2}$, 
180DWs in LNSVs with heavy-metal caplayers can preserve a fast stable steady flow with a velocity up to 
hundreds of meters per second, which provides a current-driven response with quite high efficiency 
and robustness.

\begin{figure} [htbp]
	\centering
	\includegraphics[width=0.43\textwidth]{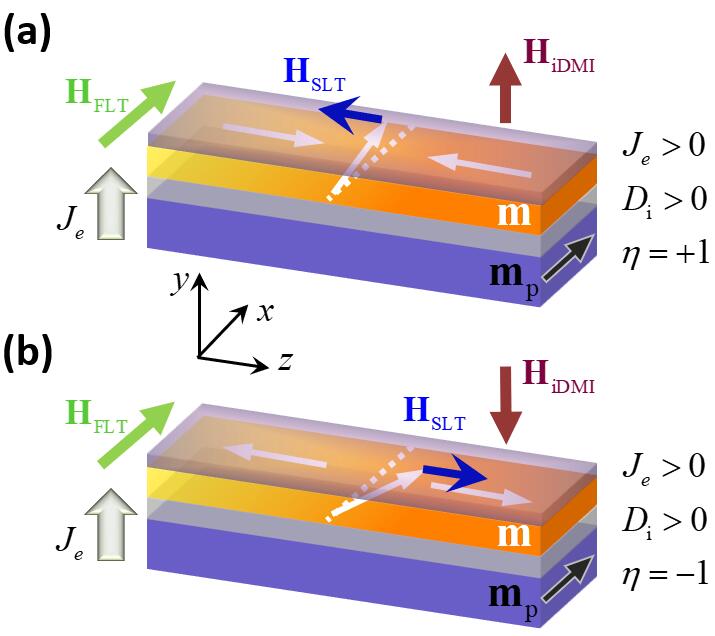}
	\caption{(Color online) Illustration of the physical mechanism responsible for the 
		``velocity saturation" phenomena at high $J_e$ under planar-transverse polarizers.
		(a) $J_e>0$, $D_{\mathrm{i}}>0$ and $\eta=+1$. 
		(b) $J_e>0$, $D_{\mathrm{i}}>0$ and $\eta=-1$. 
		In all sketches, green, blue and reddish brown arrows represent respectively
		$\mathbf{H}_{\mathrm{FLT}}$, $\mathbf{H}_{\mathrm{SLT}}$ and
		$\mathbf{H}_{\mathrm{iDMI}}$. }\label{fig3}
\end{figure}

In fact, this ``velocity saturation" phenomena can be understood physically.
Putting Eqs. (\ref{Lagrangian_density}) and (\ref{Dissipation_density})
into the Lagrangian-Rayleigh equation (\ref{Eular_Lagrangian_equation}) with $X=\theta(\phi)$, 
we obtain the familiar generalized Landau-Lifshitz-Gilbert (LLG) equation\cite{LLG_equation}
\begin{equation}\label{LLG_vector}
\begin{split}
\frac{\partial \mathbf{m}}{\partial t}=&-\gamma_0\mathbf{m}\times\mathbf{H}^0_{\mathrm{eff}}+\alpha\mathbf{m}\times\frac{\partial \mathbf{m}}{\partial t}  \\
& \quad -\gamma_0 a_J\mathbf{m}\times(\mathbf{m}\times\mathbf{m}_{\mathrm{p}})-\gamma_0 \xi a_J \mathbf{m}\times\mathbf{m}_{\mathrm{p}},
\end{split}
\end{equation}
where $\mathbf{H}^0_{\mathrm{eff}}=-(\mu_0 M_s)^{-1}\delta\mathcal{E}/\delta\mathbf{m}$.
The two effective fields related to SLT and FLT are denoted as $\mathbf{H}_{\mathrm{SLT}}=a_J(\mathbf{m}\times\mathbf{m}_{\mathrm{p}})$ and $\mathbf{H}_{\mathrm{FLT}}=\xi a_J\mathbf{m}_{\mathrm{p}}$, respectively.
For the planar-transverse polarizer $\mathbf{m}_{\mathrm{p}}=+\mathbf{e}_x$, 
$\mathbf{H}_{\mathrm{FLT}}$ is always a uniform transverse field directed along $+\mathbf{e}_x$ thus can not induce 
wall motion in $z-$direction.
However, it breaks the two-fold symmetry in $x-$direction thus the tilting angle of steady flows
tends to be parallel to it ($\varphi_2\rightarrow \phi_{\mathrm{p}}$ as $J_e$ increases).

On the other hand, the iDMI energy density leads to the following effective field
\begin{equation}\label{iDMI_effective_field}
\begin{split}
\mathbf{H}_{\mathrm{iDMI}}=&-\frac{2D_{\mathrm{i}}}{\mu_0 M_s}\left[\left(\frac{\partial m_z}{\partial z}\right)\mathbf{e}_y-\left(\frac{\partial m_y}{\partial z}\right)\mathbf{e}_z\right] \\
=&\frac{2D_{\mathrm{i}}}{\mu_0 M_s}\left[\frac{\eta\sin^2\vartheta}{\Delta}\mathbf{e}_y + \frac{\eta\sin\vartheta\cos\vartheta\sin\varphi}{\Delta}\mathbf{e}_z\right].
\end{split}
\end{equation}
Clearly the integral of $z-$component disappears thus has no effects on wall dynamics.
However, the $y-$component lifts the magnetization apart from $\phi_{\mathrm{p}}-$plane. 
In wall region the resulting $\mathbf{H}_{\mathrm{SLT}}$ points in 
$-\mathbf{e}_z(+\mathbf{e}_z)$ direction for $\eta=+1(-1)$, as shown respectively
in Fig. \ref{fig3}(a) and \ref{fig3}(b).
In both cases the wall propagates in $-\mathbf{e}_z$ direction for $D_{\mathrm{i}}>0$, 
thus explains the independence of $v_{\mathrm{sat}}$ on $\eta$ and the minus sign in 
the expression of $v_{\mathrm{sat}}$.
In addition, as current density increases the departure of wall's azimuthal plane from
$\phi_{\mathrm{p}}-$plane decreases due to the increasing $\mathbf{H}_{\mathrm{FLT}}$
under a given $D_{\mathrm{i}}>0$.
This leads to the nearly saturated $\mathbf{H}_{\mathrm{SLT}}$ thus the wall velocity
($v_{\mathrm{sat}})$.

\section{IV. 180DW dynamics under parallel polarizers}
For parallel polarizers, $\mathbf{m}_{\mathrm{p}}=\mathbf{e}_z$, 
thus $\theta_{\mathrm{p}}=0$ and $p_{\varphi}=0$. 
The dynamical equation set becomes
\begin{subequations}\label{Dynamical_equation_parallel_polarizer}
	\begin{align}
	\frac{1+\alpha^2}{\gamma_0 k_{\mathrm{H}}M_s}\frac{\eta\dot{q}}{\Delta}&=\frac{\sin 2\varphi}{2}+\frac{(1+\alpha\xi) a_J}{k_{\mathrm{H}}M_s}-\frac{\lambda}{2}\cos\varphi, \\
	\frac{1+\alpha^2}{\alpha\gamma_0 k_{\mathrm{H}}M_s}\dot{\varphi}&=\frac{\lambda}{2}\cos\varphi-\frac{\sin 2\varphi}{2}+\frac{(\xi-\alpha)a_J}{\alpha k_{\mathrm{H}}M_s}, \\
	\frac{\pi^2\alpha}{6\gamma_0 M_s}\frac{\dot{\Delta}}{\Delta}&=\frac{l_0^2}{\Delta^2}-k_{\mathrm{E}}-k_{\mathrm{H}}\sin^2\varphi.
	\end{align}
\end{subequations}
The steady-flow requirements ($\dot{\varphi}=0$ and $\dot{\Delta}=0$) lead to
\begin{equation}\label{parallel_polarizer_g_definition}
g(\varphi,\lambda)\equiv \sin 2\varphi-\lambda\cos\varphi = \frac{J_e}{J_{W,z}^{0}}\cdot\mathrm{sgn}(\xi-\alpha),
\end{equation}
where $J_{W,z}^{0}\equiv \alpha k_{\mathrm{H}}\mu_0 M_s^2 g_e d e/[2|\xi-\alpha| P\hbar]$
is the original Walker limit without iDMI ($\lambda=0$).
Eq. (\ref{parallel_polarizer_g_definition}) has the following symmetrical features ($J_e>0$):
if $\varphi_0$ is a solution for given combination of $[\lambda,J_e,\mathrm{sgn}(\xi-\alpha)]$, then:
(a) $\pi+\varphi_0$ must be a solution for $[-\lambda,J_e,\mathrm{sgn}(\xi-\alpha)]$;
(b) $\pi-\varphi_0$ must be a solution for $[\lambda,J_e,-\mathrm{sgn}(\xi-\alpha)]$.
Therefore we can focus our discussion on the case where $\lambda>0$ and $\mathrm{sgn}(\xi-\alpha)>0$.
The global maximum $G^{(0)}(\lambda)$ and minimum $-G^{(0)}(\lambda)$ of $g(\varphi,\lambda)$
take place at $\varphi=-\pi-\varphi^{(0)}$ and $\varphi=\varphi^{(0)}$, respectively.
Also, $g(\varphi,\lambda)$ has a local maximum $G^{(1)}(\lambda)$ and minimum $-G^{(1)}(\lambda)$
at $\varphi=\varphi^{(1)}$ and $\varphi=\pi-\varphi^{(1)}$, respectively.
Here
\begin{equation}\label{parallel_polarizer_g_max}
G^{(k)}(\lambda)=\sqrt{\frac{(-\lambda^4+80\lambda^2+128)+(-1)^k\lambda(\lambda^2+32)^{\frac{3}{2}}}{128}}
\end{equation}
and $\varphi^{(k)}=\sin^{-1}[(\lambda-(-1)^k\sqrt{\lambda^2+32})/8]$.
Then the iDMI-modified Walker limit is $J_{W,z}^{0}\cdot G^{(0)}(\lambda)$.
In the case of small enough $\lambda$, $G^{(0)}(\lambda)\approx 1+|\lambda|/\sqrt{2}$
indicating the enhancement of Walker limit by iDMI.

When $0<J_e/J_{W,z}^{0}<G^{(1)}(\lambda)[G^{(1)}(\lambda)<J_e/J_{W,z}^{0}<G^{(0)}(\lambda)]$, 
four (two) solutions emerges.
We then perform stability analysis for these steady flows.
For variation of $\varphi=\varphi_i+\delta\varphi_i$, Eq. (\ref{Dynamical_equation_parallel_polarizer}b) provides
\begin{equation}\label{Stability_analysis_phi_parallel_polarizer}
\frac{1+\alpha^2}{\alpha\gamma_0 k_{\mathrm{H}}M_s}\frac{\partial(\delta\varphi_i)}{\partial t}=-\frac{1}{2}\left[\frac{\partial g(\varphi,\lambda)}{\partial \varphi}\right]_{\varphi=\varphi_i}\delta\varphi_i.
\end{equation}
Therefore, solutions on the ``upper slopes" are stable.
On the other hand, for variation of wall width: $\Delta=\Delta(\varphi_i)+\delta\Delta_i$, one has
\begin{equation}\label{Stability_analysis_Delta_parallel_polarizer}
\frac{\pi^2\alpha}{6\gamma_0 M_s}\frac{\partial(\delta\Delta_i)}{\partial t}=-\frac{2l_0^2}{\Delta^2(\varphi_i)}\delta\Delta_i,
\end{equation}
which ascertains the persistent stability of wall width.
At last, in stable steady flows Eq. (\ref{Dynamical_equation_parallel_polarizer}) provides the wall velocity
as $\eta\dot{q}=\xi\gamma_0 a_J\Delta(\varphi_i)/\alpha$, where the effect of iDMI resides in the 
modification to wall width.
In the ``weak iDMI" limit ($|\lambda|\ll 1$), the relative change of wall velocity reads
\begin{equation}\label{parallel_polarizer_steady_v_change_small_iDMI}
\frac{\dot{q}(\lambda)-\dot{q}(0)}{\dot{q}(0)}=-\frac{k_{\mathrm{H}}\sin 2\varphi_i\cos\varphi_i}{4(k_{\mathrm{E}}+k_{\mathrm{H}}\sin^2\varphi_i)\cos 2\varphi_i}\lambda\sim O(\lambda),
\end{equation}
where $\varphi_i$ is the corresponding stable solution of Eq. (\ref{parallel_polarizer_g_definition}).

When $J_e/J_{W,z}^{0}>G^{(0)}(\lambda)$, the wall begins to precess.
For a physical quantity $O$, its time average 
$\langle O\rangle\equiv T^{-1}\int_{0}^{T}X(t)\mathrm{d}t=T^{-1}\int_{0}^{2\pi}(X/\dot{\varphi})\mathrm{d}\varphi$
corresponds to experimental observables, where $T$ is the precession period.
In the absence of iDMI, $T=4\pi(1+\alpha^2) J_{W,z}^0/[\alpha\gamma_0 k_{\mathrm{H}} M_s\sqrt{(J_e)^2-(J_{W,z}^0)^2}]$.
When iDMI emerges, in principle $T$ can not be analytically integrated out.
In the limit of weak iDMI ($|\lambda|\ll 1$) and high current ($J_e/J_{W,z}^{0}\gg 1$), 
we preserve the form of $T$ but replace $J_{W,z}^0$ by $J_{W,z}^{0}\cdot G^{(0)}(\lambda)$.
To the linear terms of $\lambda$ and $J_e/J_{W,z}^{0}$, the new period $T'$ reads
$T'=T\cdot ( 1+|\lambda|/\sqrt{2})$.
Finally the time-averaged wall velocity takes the following form:
\begin{equation}\label{parallel_polarizer_precessional_v_small_iDMI}
\eta \langle \dot{q}\rangle=\frac{1+\alpha\xi}{1+\alpha^2}\gamma_0 a_J\Delta + \frac{\pi\Delta}{\alpha T}\left(1-\frac{|\lambda|}{\sqrt{2}}\right)\left(\frac{J_{W,z}^0}{J_e}\right)^2,
\end{equation}
in which $\Delta$ and $T$ are the time-averaged wall width and precession period in the absence of iDMI, respectively.
Clearly, wall velocities in steady and precessional flows both depend on the topological wall charge $\eta$.

\section{V. 180DW dynamics under perpendicular polarizers}
For perpendicular polarizers, $\mathbf{m}_{\mathrm{p}}=\mathbf{e}_y$. 
Consequently, Eq. (\ref{Dynamical_equation_original}) is simplified to
\begin{subequations}\label{Dynamical_equation_perp_polarizer}
	\begin{align}
	\frac{1+\alpha^2}{\gamma_0 k_{\mathrm{H}}M_s}\frac{\eta\dot{q}}{\Delta}&=\left[\sin\varphi+\frac{\pi}{2}\frac{(\alpha -\xi)a_J}{k_{\mathrm{H}}M_s}-\frac{\lambda}{2}\right]\cos\varphi, \\
	\frac{1+\alpha^2}{\alpha \gamma_0  k_{\mathrm{H}} M_s}\dot{\varphi}&=\left[\frac{\pi}{2}\frac{(1+\alpha\xi)a_J}{\alpha  k_{\mathrm{H}} M_s} -\sin\varphi + \frac{\lambda}{2}\right]\cos\varphi, \\
	\frac{\pi^2\alpha}{6\gamma_0 M_s}\frac{\dot{\Delta}}{\Delta}&=\left(\frac{l_0^2}{\Delta^2}-k_{\mathrm{E}}-k_{\mathrm{H}}\sin^2\varphi\right)+\frac{\pi \xi a_J}{M_s}\sin\varphi.
	\end{align}
\end{subequations}
For steady flows, $\dot{\varphi}=0$ and $\dot{\Delta}=0$, thus lead to two branches of solution.
The first one is a stationary rigid-body
\begin{eqnarray}\label{Solution_branch_1_perp_polarizer}
&\varphi=\left(n+\frac{1}{2}\right)\pi,\quad \dot{q}=0,&  \nonumber  \\ &\Delta(\varphi)=l_0\left[k_{\mathrm{E}}+k_{\mathrm{H}}-(-1)^n\frac{\pi \xi a_J}{M_s}\right]^{-1/2},&
\end{eqnarray}
and the second is an ordinary steady flow
\begin{subequations}\label{Solution_branch_2_perp_polarizer}
\begin{align}
u(\varphi',\lambda)&\equiv\sin\varphi'-\frac{\lambda}{2}=\frac{J_e}{J_{W,x}^0}>0, \\
\dot{q}'&=\frac{\pi}{2}\frac{\eta\Delta(\varphi'_0)\gamma_0 a_J}{\alpha}\cos\varphi',    \\
\Delta(\varphi')&=l_0\left(k_{\mathrm{E}}+k_{\mathrm{H}}\sin^2\varphi'-\frac{\pi \xi a_J}{M_s}\sin\varphi'\right)^{-1/2}.
\end{align}
\end{subequations}

We first determine the modified Walker limit of steady flow.
Since $J_e$ is always positive, $|\lambda|<2$ is a convenient constraint that ensures 
the existence of solutions to Eq. (\ref{Solution_branch_2_perp_polarizer}a).
Obviously, $u(\varphi',\lambda)$ acquires its maximum $1-\lambda/2$ at $\varphi'=\pi/2$,
leading to the new Walker limit as $J_{W,x}^0\cdot(1-\lambda/2)$.
However, at this very point the wall velocity becomes zero thus merges with the other stationary branch.

When $0<J_e/J_{W,x}^0<1-\lambda/2$, there are always two options of steady flow. 
We then perform stability analysis to testify their robustness.
Putting the variance $\varphi=\varphi'+\delta\varphi'$ into Eq. (\ref{Dynamical_equation_perp_polarizer}b),
we find 
\begin{equation}\label{Stability_analysis_phiprime_perp_polarizer}
\frac{\partial(\delta\varphi')}{\partial t}=-\frac{\alpha\gamma_0 M_s k_{\mathrm{H}}\cos^2\varphi'}{1+\alpha^2}\delta\varphi',
\end{equation}
implying that both are stable. The corresponding wall velocity reads
\begin{equation}\label{TDW_velocity_branch_2_perp_polarizer}
\eta\dot{q}'=\frac{\pi}{2}\frac{l_0\gamma_0 a_J}{\alpha}\frac{\cos\varphi'}{\sqrt{k_{\mathrm{E}}+k_{\mathrm{H}}\sin^2\varphi'-\frac{\pi \xi a_J}{M_s}\sin\varphi'}},
\end{equation}
with $\sin\varphi'=(\lambda/2)+(J_e/J_{W,x}^0)$.
Obviously, these two steady flows have symmetrical titling angles about $\pi/2\cdot\mathrm{sgn}(\sin\varphi')$, 
the same wall width and opposite velocities.

Finally we go back to the stationary mode in Eq. (\ref{Solution_branch_1_perp_polarizer}). 
By introducing similar variance $\varphi+\delta\varphi$ and performing standard algebra, we get
\begin{equation}\label{Stability_analysis_phi_perp_polarizer}
\frac{1+\alpha^2}{\alpha\gamma_0 k_{\mathrm{H}}M_s}\frac{\partial(\delta\varphi)}{\partial t}=1-(-1)^n\left(\frac{J_e}{J_{W,x}^0}+\frac{\lambda}{2}\right).
\end{equation}
Obviously, this stationary mode will be stable when and only when $n$ is even and $J_e$ exceeds 
the modified Walker limit $J_{W,x}^0\cdot(1-\lambda/2)$.
This directly results in the absence of precessional flow under perpendicular polarizers.
This conclusion also holds for large enough iDMI ($|\lambda|>2$).

\section{VI. Disscussions}
First, in all our deductions we introduce a dimensionless variable 
$\lambda=\eta D_{\mathrm{i}}\pi/(k_{\mathrm{H}}\mu_0 M_s^2\Delta)$
to account for the effects of iDMI.
Generally the wall width $\Delta$ is also modified by iDMI.
However, one can prove by series expansions that the correction to $\lambda$ by wall width
is a second-order term thus can be neglected when $|\lambda|\ll 1$.
Therefore, the introduction of $\lambda$ is reasonable as long as
the iDMI is not too high which is common in real spin valves.

Second, we would like to emphasize again that the disappearance mechanisms of 
precessional flows under planar-transverse and perpendicular polarizers are different.
In the former case, the Walker breakdown is postponed significantly due to the 
consistent existence of $\varphi_2-$solution (stable steady flow) in Fig. \ref{fig2} as the current
density increases. 
Any other tilting angle of a 180DW will finally fall into $\varphi_2$ meantime 
bearing a finite wall velocity which converges to a saturation value shown in 
Eq. (\ref{planar_transverse_v_saturation}).
However, in the latter case the corresponding Walker limit is only modulated to a finite extent.
When current density is below this critical value, two branches of steady flow can be solved 
out in which the stationary one is unstable.
Interestingly when current density exceeds the new Walker limit, this stationary mode
becomes stable. A 180DW with any tilting angle will eventually fall into this ``potential valley"
and then keeps static.
Of course, all our discussions depend on the existence of 180DWs with the generalized Walker
profile in Eq. (\ref{Walker_static_generalized}).
Under high enough current densities, this ideal profile 
could be destroyed by the twisting SLTs across the wall region
and 180DWs would change to other types of magnetic soliton or even collapse and submerge
into strong spin waves. This would greatly complicate the analytics thus beyond the
scope of this work.

At last, for free layers with PMA, parallel investigations have been performed. 
After surveying the results in both IPMA and PMA cases, 
in the presence of iDMI the following general correspondence can be summarized:
(i) when the polarizer is parallel to the easy axis [i.e. parallel (perpendicular) polarizer 
for IPMA (PMA)], the Walker limit is enlarged by a factor $G^{(0)}(\lambda)$
meantime both steady and precessional flows survive. 
(ii) when the polarizer is perpendicular to the easy axis but still in $yz-$plane 
[i.e. perpendicular (parallel) polarizer for IPMA (PMA)], 
the Walker limit is corrected by an additional factor 
[$1-\lambda/2$ for IPMA and $1+\lambda/2$ for PMA]
and the precessional flow vanishes.
(iii) under planer-transverse polarizers for both IPMA and PMA cases, 
the Walker breakdown is postponed considerably and the surviving steady flow
is stabilized by iDMI with an iDMI-determined saturation velocity.

\section{\label{Section_Conclusion} VII. Summary}
In this work, 180DW dynamics in LNSVs with heavy-metal caplayers under perpendicularly injected currents are systematically investigated. 
Both IPMA and PMA in free layers of LNSVs are considered.
Since the results are quite similar, the IPMA case is taken as an example to illustrate our main findings.
Surprisingly, under planar-transverse polarizers the Walker breakdown is pushed away considerably
by iDMI. More importantly, the originally unstable steady flows of 180DWs are stabilized by iDMI
and acquire high saturation velocity solely determined by iDMI strength under not-too-large 
current densities.
For parallel polarizers, the modifications of iDMI to Walker limit, as well as wall velocity in both steady and 
precessional flows are provided.
At last, for perpendicular polarizers we reveal that precessional flows disappear due to the
stable stationary mode above the new Walker limit.
These results provide insights for developing magnetic nanodevices with low energy consumption
and high robustness.


\section{Acknowledgement}
M. L. is supported by the National Natural Science Foundation of China (Grant No. 11947023),
the Project of Hebei Province Higher Educational Science and Technology Program (QN2019309)
and the PhD Research Startup Foundation of Shijiazhuang University (20BS022).
J. L. acknowledges support from Natural Science Foundation for Distinguished Young Scholars of Hebei Province of China (A2019205310).


\begin{thebibliography}{999}

\bibitem{Slonczewski_JMMM_1996} J. Slonczewski, J. Magn. Magn. Mater. \textbf{159}, L1 (1996).



\bibitem{Fert_JAP_2002} J. Grollier, D. Lacour, V. Cros, A. Hamzic, A. Vaur\`{e}s, A. Fert, D. Adam and G. Faini, J. Appl. Phys. \textbf{92}, 4825 (2002).
\bibitem{Fert_APL_2003} J. Grollier, P. Boulenc, V. Cros, A. Hamzi\'{c}, A. Vaur\`{e}s, A. Fert, and G. Faini, Appl. Phys. Lett. \textbf{83}, 509 (2003). 
\bibitem{Lim_APL_2004} C. K. Lim, T. Devolder, C. Chappert, J. Grollier, V. Cros, A. Vaur\`{e}s, A. Fert, and G. Faini, Appl. Phys. Lett. \textbf{84}, 2820 (2004). 

\bibitem{Rebei_Mryasov_PRB_2006} A. Rebei and O. Mryasov, Phys. Rev. B \textbf{74}, 014412 (2006).
\bibitem{Kawabata_IEEE_2011} K. Kawabata, M. Tanizawa, K. Ishikawa, Y. Inoue, M. Inuishi, and T. Nishimura, in \textit{2011 International Conference on Simulation of Semiconductor Processes and Devices, 8-10 September 2011, Osaka, Japan} (IEEE, Piscataway, NJ, 2011), pp. 55–58.

\bibitem{Khvalkovskiy_PRL_2009} A. V. Khvalkovskiy, K. A. Zvezdin, Ya. V. Gorbunov, V. Cros, J. Grollier, A. Fert, and A.K. Zvezdin, Phys. Rev. Lett. \textbf{102}, 067206 (2009).

\bibitem{Boone_PRL_2010_exp} C. T. Boone, J. A. Katine, M. Carey, J. R. Childress, X. Cheng, and I. N. Krivorotov, Phys. Rev. Lett. \textbf{104}, 097203 (2010).

\bibitem{Grollier_NatPhys_2011} A. Chanthbouala, R. Matsumoto, J. Grollier, V. Cros, A. Anane, A. Fert, A. V. Khvalkovskiy, K. A. Zvezdin, K. Nishimura, Y. Nagamine, H. Maehara, K. Tsunekawa, A. Fukushima, and S. Yuasa, Nat. Phys. \textbf{7}, 626 (2011).
\bibitem{Metaxas_SciRep_2013} P. J. Metaxas, J. Sampaio, A. Chanthbouala, R. Matsumoto, A. Anane, A. Fert, K. A. Zvezdin, K. Yakushiji, H. Kubota, A. Fukushima, S. Yuasa, K. Nishimura, Y. Nagamine, H. Maehara, K. Tsunekawa, V. Cros, and J. Grollier, Sci. Rep. \textbf{3}, 1829 (2013).
\bibitem{Grollier_APL_2013}J. Sampaio, S. Lequeux, P. J. Metaxas, A. Chanthbouala, R. Matsumoto, K. Yakushiji, H. Kubota, A. Fukushima, S. Yuasa, K. Nishimura, Y. Nagamine, H. Maehara, K. Tsunekawa, V. Cros, and J. Grollier, Appl. Phys. Lett. \textbf{103}, 242415 (2013).

\bibitem{jlu_PRB_2019} M. Li, Z. An, and J. Lu, Phys. Rev. B \textbf{100}, 064406 (2019).

\bibitem{He_EPJB_2013} P.-B. He, Eur. Phys. J. B \textbf{86}, 412 (2013).

\bibitem{Boulle_PRL_2013} O. Boulle, S. Rohart, L. D. Buda-Prejbeanu, E. Ju\'{e}, I. M. Miron, S. Pizzini, J. Vogel, G. Gaudin, and A. Thiaville, Phys. Rev. Lett. \textbf{111}, 217203 (2013).

\bibitem{Dzyaloshinsky} I. Dzyaloshinsky, J. Phys. Chem. Solids \textbf{4}, 241 (1958). 
\bibitem{Moriya} T. Moriya, Phys. Rev. \textbf{120}, 91 (1960)


\bibitem{Bogdanov_JMMM_1994} A. Bogdanov and A. Hubert, J. Magn. Magn. Mater. \textbf{138}, 255 (1994).


\bibitem{Aharoni_JAP_1998} A. Aharoni, J. Appl. Phys. \textbf{83}, 3432 (1998).
\bibitem{jlu_PRB_2016} J. Lu, Phys. Rev. B \textbf{93}, 224406 (2016).
\bibitem{jlu_SciRep_2017} M. Li, J. B. Wang, and J. Lu, Sci. Rep. \textbf{7}, 43065 (2017).
\bibitem{jlu_PRB_2020} J. Lu, M. Li, and X. R. Wang, Phys. Rev. B \textbf{101}, 134431 (2020).
\bibitem{jlu_JMMM_2021} M. Li and J. Lu, J. Magn. Magn. Mater. \textbf{525}, 167684 (2021).


\bibitem{Lee_PhysRep_2013} K.-J. Lee, M. D. Stiles, H.-W. Lee, J.-H. Moon, K.-W. Kim, and S.-W. Lee, Phys. Rep. \textbf{531}, 89 (2013).
\bibitem{Chshiev_PRB_2015} M. Chshiev, A. Manchon, A. Kalitsov, N. Ryzhanova, A. Vedyayev, N. Strelkov, W. H. Butler, and B. Dieny, Phys. Rev. B \textbf{92}, 104422 (2015).


\bibitem{Walker_JAP_1974} N. L. Schryer and L. R. Walker, J. Appl. Phys. \textbf{45}, 5406 (1974).


\bibitem{LLG_equation} T. L. Gilbert, IEEE Trans. Magn. \textbf{40}, 3443 (2004).










	
\end{thebibliography}

\end{document}